\documentclass[conference]{IEEEtran}
\def\IEEEtitletopspace{18pt}
\usepackage{soul}
\usepackage{cite}
\usepackage{amsmath,amssymb,amsfonts,mathtools}
\usepackage{algorithmic}
\usepackage{graphicx}
\usepackage{textcomp}
\usepackage{xcolor}
\usepackage{multirow}
\usepackage{subfigure}
\usepackage[symbol]{footmisc}
\def\BibTeX{{\rm B\kern-.05em{\sc i\kern-.025em b}\kern-.08em
    T\kern-.1667em\lower.7ex\hbox{E}\kern-.125emX}}

\DeclareMathOperator{\tr}{tr}
\DeclareMathOperator{\E}{E}
\DeclareMathOperator{\A}{A}
\DeclareMathOperator{\B}{B}
\DeclareMathOperator{\var}{cov}
\DeclareMathOperator{\Exp}{Exp}

\begin{document}

\title{Adaptive Invariant Extended Kalman Filter with Noise Covariance Tuning for Attitude Estimation}

\author{\IEEEauthorblockN{Yash Pandey\IEEEauthorrefmark{1}}
\IEEEauthorblockA{\textit{Department of Electronics Engineering} \\
\textit{Harcourt Butler Technical University}\\
Kanpur, India \\
yashpanknp@gmail.com}

\and
\IEEEauthorblockN{Rahul Bhattacharyya\IEEEauthorrefmark{1}}
\IEEEauthorblockA{\textit{Department of Electrical Engineering} \\
\textit{Indian Institute of Technology, Kanpur} \\
Kanpur, India \\
rahulbhatta0@gmail.com}

\and
\IEEEauthorblockN{Yatindra Nath Singh}
\IEEEauthorblockA{\textit{Department of Electrical Engineering} \\
\textit{Indian Institute of Technology, Kanpur} \\
Kanpur, India \\
ynsingh@iitk.ac.in}
}
\maketitle
\footnotetext[1]{Contributed Equally}

\begin{abstract}
Attitude estimation is crucial in aerospace engineering, robotics, and virtual reality applications, but faces difficulties due to nonlinear system dynamics and sensor limitations. This paper addresses the challenge of attitude estimation using quaternion-based adaptive right invariant extended Kalman filtering (RI-EKF) that integrates data from inertial and magnetometer sensors. Our approach applies the expectation-maximization (EM) algorithm to estimate noise covariance, exploiting RI-EKF symmetry properties. We analyze the adaptive RI-EKF's stability, convergence, and accuracy, validating its performance through simulations and comparison with the left invariant EKF. Monte Carlo simulations validate the effectiveness of our noise covariance estimation technique across various window lengths.
\end{abstract}


\section{Introduction}
Estimating the orientation of an object relative to a reference frame, commonly referred to as attitude estimation, is a significant challenge in fields such as aerospace engineering, robotics, and virtual reality \cite{schmidt1981,moore2016}. Accurate orientation information is critical for numerous applications, ranging from aircraft or UAV control, navigation sensors in smartphones, and headsets for augmented and virtual reality applications\cite{groves2015}. The primary difficulty lies in reliably sustained determination of an object’s orientation with respect to an accepted frame of reference in three-dimensional space, especially when the object is dynamic in the reference frame \cite{woodman2007}. \par
Modern attitude estimation systems typically rely on a combination of inertial sensors (gyroscopes and accelerometers) and magnetic sensors (magnetometers) to continuously estimate the change in orientation with respect to the initial known orientation \cite{titterton2004}. Each sensor type presents unique challenges. Gyroscopes measure angular velocity but suffer from bias and drift over time, leading to accumulated errors in orientation estimates. Accelerometers can provide information about the gravity vector but are also sensitive to linear accelerations of the body, making it difficult to distinguish between gravitational and inertial accelerations due to movements. Magnetometers offer absolute heading information by measuring the Earth’s magnetic field but are susceptible to local magnetic disturbances and variations in the Earth's field. Furthermore, all these sensors are subject to various sources of noise, including thermal noise, quantization errors, and environmental factors. The integration of data from these diverse sensors, each with its error characteristics, poses a significant challenge in achieving robust and accurate attitude estimates \cite{quinchia2013}.\par
The Kalman filter, introduced by Rudolf E. Kálmán in 1960 \cite{kalman1960}, has become a foundation in estimation theory and has found wide application in attitude estimation. The basic Kalman filter provides an optimal estimate for linear systems with Gaussian noise. However, attitude estimation is inherently nonlinear due to the rotational nature of orientation, necessitating the use of nonlinear variants of the Kalman filter \cite{crassidis2007}. The extended Kalman filter (EKF) has been a popular choice for nonlinear estimation problems, including attitude estimation. It works by linearizing the nonlinear system around the current estimate. While effective in many scenarios, the EKF can suffer from linearization errors and may lead to inconsistent forecasts, especially in highly nonlinear systems or when initial estimates are poor. Other variants, such as the unscented Kalman filter (UKF) and particle filters, have been developed to address some of the EKF’s limitations \cite{simon_ekf}. These methods aim to capture the nonlinear behavior of the system better but often come at the cost of increased computational complexity.
\par In recent years, the application of invariant theory to estimation problems has gained traction, leading to the development of the right invariant extended Kalman filter (RI-EKF) \cite{bonnabel2009,gui2018}. The key idea behind invariant filtering is to exploit the underlying geometric structure of the state space, particularly its symmetries \cite{barrau2018,barrau2023}. For attitude estimation, the state space is the special orthogonal group $SO(3)$ or its double cover, the unit-quaternion. These spaces possess rich geometric properties, including group structure and invariances under certain transformations. By designing filters that respect these invariances, it’s possible to achieve estimates that are more consistent and robust to linearization errors.
The RI-EKF formulates the estimation problem in a way that preserves the intrinsic invariances of the system. This approach has shown promising results in navigation and attitude estimation tasks, often outperforming traditional EKF implementations in terms of consistency and convergence properties.
Despite the advancements in attitude estimation techniques, several challenges remain, particularly in adaptive filtering for systems with varying noise characteristics \cite{mehra1970,mohamed1999,shumway1982,ananthasayanam2016}. \par This research addresses a critical gap in the field by focusing on the following problem. How can we develop an adaptive RI-EKF for quaternion-based attitude estimation that effectively handles varying sensor noise characteristics while maintaining the geometric advantages of invariant filtering?
\par The main contributions of this article are:
\begin{itemize}
\item To analyze the theoretical properties of the proposed adaptive RI-EKF in terms of stability, convergence, and estimation accuracy.
\item Applying the EM algorithm for estimation of noise covariance by exploiting the symmetry of the RI-EKF 
\item To validate the performance of the adaptive RI-EKF through comprehensive simulations and comparision with the LI-EKF.
\end{itemize}

\section{Background}

\subsection{Quaternions and Rotation}
The unit-quaternion is a popular parameterization of rotation, more efficient than directional cosine matrices (DCM) and more robust than Euler angles \cite{shoemake1985}. Being isomorphic to the 3-sphere $S^3$, it is a double cover to the three-dimensional rotation group $SO(3)$. With $\mathbb{H} \coloneq \left\{\begin{bmatrix}
    q_w & q_v^T
\end{bmatrix}^T\mid q_w\in\mathbb{R},q_v\in\mathbb{R}^3\right\}$ as the set of all quaternions, we write 
\begin{equation}
    S^3 = \left\{q\in\mathbb{H} \mid \|q\|=1\right\},
\end{equation}
where we define the norm as $\|q\|^2 \coloneq q\otimes q^*$ and the conjugate as $q^* \coloneq \begin{bmatrix}
    q_w & -q_v^T
\end{bmatrix}^T$. The quaternion multiplication $\otimes$ is defined as
\begin{align*}
    p\otimes q &= \begin{bmatrix}
        p_w q_w - p_v^T q_v \\
        p_w q_v + p_v q_w + \left[p_v\right]_\times q_v 
    \end{bmatrix} \\
    &= \Xi[p] q = \Omega[q] p,
\end{align*}
with the matrices $\Xi$ and $\Omega$ encoding quaternion multiplication as matrix multiplication, given by
\begin{align*}
    \Xi[p] &= \begin{bmatrix}
        p_w & -p_v^T \\ p_v & p_wI_3 + [p_v]_\times
    \end{bmatrix}, \\
    \Omega[q] &= \begin{bmatrix}
        q_w & -q_v^T \\ q_v & q_wI_3 - [q_v]_\times
    \end{bmatrix},
\end{align*}
and $[v]_\times$, for $v=\begin{bmatrix}
    v_x & v_y & v_z
\end{bmatrix}$, is the skew-symmetric matrix
\begin{equation*}
    \begin{bmatrix}
    0 & -v_z & v_y \\
    v_z & 0 & -v_x \\
    -v_y & v_x &0
\end{bmatrix}.
\end{equation*}
Further, quaternion-vector multiplication is defined as
$q\otimes v \coloneq q \otimes \begin{bmatrix}
    0 & v^T
\end{bmatrix}^T$. 

Rotation of a vector $r\in \mathbb{R}^3$ by a quaternion $q \in S^3$ is given by 
\begin{align}
\label{quat_rot}
    r_q &= q^*\otimes r\otimes q \nonumber \\
    &= \A(q) r,
\end{align}
with $\A$ being the map $S^3 \to SO(3)$ from the unit-quaternion to the DCM representation corresponding to a certain rotation:
\begin{equation}
    \A: q \mapsto (q_w^2 - q_v^Tq_v)I + 2q_vq_v^T - 2q_w\left[q_v\right]_\times
.\end{equation}
Note that $\A$ is surjective but not injective. In fact, $\A\left(q\right) = \A\left(-q\right)$ for $q\in S^3$, i.e., the unit-quaternions $q$ and $-q$ represent equivalent rotations (this is the double cover property of $S^3$). In the convention that follows, equation \eqref{quat_rot} represents the rotation of the vector $v$ in the global/inertial frame to the local/body frame. 

We also define an exponential map $\mathbb{R}^3 \to S^3$ as \begin{equation}
	\label{exp_def}
	\Exp: \xi \mapsto \begin{bmatrix} \cos\|\xi\| \\ \frac{\xi}{\|\xi\|} \sin\|\xi\| \end{bmatrix} 
.\end{equation}  
Note that $\Exp\left( \xi \right) \to \begin{bmatrix} 1 & \xi^T \end{bmatrix}^T$ as $\xi \to \begin{bmatrix} 0&0&0 \end{bmatrix}^T$.
\subsection{Attitude Estimation}
Based on the sensor models for gyroscope,  accelerometer and magnetometer, the attitude estimation problem can be formulated as \cite{sola2017,crassidis2007}
\begin{align}
	\label{noiseless_prop}
	\dot{q} &= \frac{1}{2} q \otimes \omega, \\
	\label{noiseless_meas}
	z_k &= \begin{bmatrix} a_k \\ m_k \end{bmatrix}  =  \begin{bmatrix} A(q_k)^Tg \\ A(q_k)^Tb \end{bmatrix}
,\end{align}
where $q\in S^3$ represents the orientation of the body, $\omega\in\mathbb{R}^3$ is its angular velocity (measured by the gyroscope), $a\in\mathbb{R}^3$ is the true acceleration (measured by the accelerometer) and $m\in\mathbb{R}^3$ is the magnetic field (measured by the magnetometer). $a$ and $m$ are the earth's gravitational field $g$ and magnetic field $b$ respectively, rotated to the body frame.

In practice, these measurements are corrupted by noise, assumed to be zero-mean Gaussian noise:
\begin{align*}
    \bar{\omega} &= \omega + \eta, \\
    \bar{a}_k &= a_k + \nu^a_k, \quad
    \bar{m}_k = m_k + \nu^m_k,
\end{align*}
where $\bar{\omega},\bar{a}_k,\bar{m}_k\in\mathbb{R}^3$ denote the corrupted measurements, with the noises $\eta$, $\nu^a$ and $\nu^m$ given by

\begin{align}
    \label{gyro_noise}
    \eta&\sim\mathcal{N}(0, \Sigma_\eta), \\
    \label{meas_noise}
    \nu_k^a&\sim\mathcal{N}(0, \Sigma_a), \quad
    \nu_k^m\sim\mathcal{N}(0, \Sigma_m). \nonumber
    \end{align}
Defining 
\begin{align}
    \nu_k=\begin{bmatrix}
    \nu_k^a \\ \nu_k^m
\end{bmatrix}&\sim\mathcal{N}(0, \Sigma_\nu), \quad \Sigma_\nu = \begin{bmatrix}
        \Sigma_a & 0 \\
        0 & \Sigma_m
    \end{bmatrix},
\end{align} 
the complete model can be written as
\begin{align}
	\label{true_prop}
	\dot{q} &= \frac{1}{2} q \otimes \left( \bar{\omega} - \eta \right), \\
	\label{true_meas}
	\bar{z}_k &= \begin{bmatrix} a_k \\ m_k \end{bmatrix} + \begin{bmatrix} \nu_k^a \\ \nu_k^m \end{bmatrix} = z_k  + \nu_k
.\end{align}
Equations \eqref{true_prop} and \eqref{true_meas} can be used directly as the propagation and measurement models for the EKF, with the quaternion $q$ as the state. However, this poses several theoretical problems.

The Kalman filter is generally defined for vector-states that compose over addition, i.e., if $x_1$ and $x_2$ are valid states, then so is $x_1 + x_2$. However, this is not true of our particular representation of orientation - the unit-quaternion. In fact, addition violates the unit-norm constraint: $\|x_1 + x_2\| \geq \|x_1\| + \|x_2\|$. One can verify from equation \eqref{quat_rot} that if $q_1$ and $q_2$ are two individual rotations such that $q_1,q_2\in S^3$, then $q_1\otimes q_2$ represents the combined rotation, not $q_1 + q_2$. It is obvious then that the update step of the Kalman filter, which acts on vector spaces to (additively) produce the posterior estimate will not produce a unit-norm quaternion \cite{pandey2024}. In practice, a brute-force approach of renormalizing the quaternion after every filter iteration is often used to remedy this problem. While this produces somewhat satisfactory results, the conceptual flaw remains and improvements can be made \cite{markley2004}.

Further, the non-linearity in the second term in equation \eqref{true_prop} ensures that $q$ is not a Gaussian. Indeed, if $q$ was a Gaussian, the unit-norm constrained would once again be violated \cite{markley2004}.

The EKF deals with the non-linearity in \eqref{true_prop} by linearizing the system around the present estimate, which introduces a state-dependence on the noise dynamics. Many noise covariance estimation algorithms require a constant true covariance, making this model unsuitable for adaptive estimation.

\section{Adaptive RI-EKF}

\subsection{Right-Invariant Extended Kalman Filter}
The conceptual problems described in the previous section motivate the construction of a filter that preserves the structural properties, i.e. the unit-norm nature, of the quaternion representation of orientation. Kalman filters on manifolds $SO(3)$ (rotation matrices) and $S^3$ (unit-quaternions) have been proposed in \cite{markley2004,bonnabel2009,bourmaud2015,bourmaud2013,hauberg2013,gui2018,hartley2020,pandey2024}. One such filter is the invariant extended Kalman filter \cite{barrau2016,barrau2018}, whose right invariant variety is particularly well-suited for the attitude-estimation problem \cite{gui2018,hartley2020}. 

Defining the right invariant error for the true quaternion $q$ and the estimated quaternion $\hat{q}$ as 
\begin{equation}
	\label{rierr_def}
	\varepsilon \coloneq \hat{q} \otimes q^{-1} = \hat{q} \otimes \bar{q}
\end{equation}
$q$ is defined by equation \eqref{true_prop} and $\hat{q}$ is  given by the noiseless dynamics of \eqref{noiseless_prop}:
\[
\dot{\hat{q}} = \frac{1}{2} \hat{q} \otimes \omega
.\] 
The dynamics of $\varepsilon$ can thus be written as \begin{align}
	\label{err_prop}
	\dot{\varepsilon} &= \dot{\hat{q}} \otimes q^{-1} + \hat{q} \otimes \dot{q}^{-1} \nonumber \\
			  &=  \frac{1}{2} \hat{q} \otimes \eta \otimes q^{-1} \nonumber \\
			  &= \frac{1}{2} \left[ \hat{q} \otimes \eta \otimes \hat{q}^{-1} \right] \otimes \varepsilon = \frac{1}{2} \hat{\eta} \otimes \varepsilon
.\end{align} 
Writing $\varepsilon = \Exp\left( \xi / 2 \right) \approx \begin{bmatrix} 1 & \xi^T / 2 \end{bmatrix}^T$ for some small $\xi \in \mathbb{R}^3$ and expanding equation \eqref{err_prop} while only preserving the first order terms of $\xi$ and $\eta$, we get 
\begin{align}
	\label{xi_diff}
	\dot{\xi} &= \hat{\eta} + \frac{1}{2} \left[ \hat{\eta} \right]_\times \xi, \nonumber \\
	&\approx \hat{\eta}
.\end{align}
Discretizing equation \eqref{xi_diff} with timestep $\Delta t$,
\begin{equation}
	\label{xi_prop}
	\xi_{k+1} = \xi_k + \hat{\eta}_k \Delta t
,\end{equation}
with $\hat{\eta}_k = A(\hat{q}_k) \eta_k$. 
We have reduced the dynamics of the unit-quaternion, which is constrained to lie on $S^3$, to the dynamics of the tangent $\xi$, lying in $\mathbb{R}^3$, which has no structural constraint and can be used as the state for the filter. 

Note that the noiseless dynamics (with $\eta$ set to zero) described by equations \eqref{xi_diff} and \eqref{xi_prop} are completely independent of the orientation trajectory: $\dot{\xi} = 0$, $\xi_{k+1} = \xi_k$. As opposed to the left invariant variety \cite{pandey2024,gui2018}, which formulates the dynamics of the estimation error in the body frame, the RI-EKF considers these dynamics in the inertial frame, leading to the state-independence property. While not presented here, this property is especially useful in more sophisticated applications where additional states need to monitored: bias compensation \cite{gui2018}, velocity-aided attitude estimation \cite{bonnabel2009} and navigation and simultaneous localization and mapping (SLAM) problems \cite{hartley2020}. 

The gyroscope noise $\eta$ however, is attached to the body frame and must be rotated into the global frame. This is the role of the $A(\hat{q}_k)$ term in the expression of $\hat{\eta}$. A special case exists if the noise $\eta$ is isotropic, i.e. $\Sigma_\eta = \sigma_\eta^2 I$ for $\sigma_\eta \in \mathbb{R}$, its behavior (as a Gaussian) does not change with rotation (it is obvious that the covariance of $\eta$ and $\hat{\eta}$ are the same for such a noise) and hence the rotation term $A(\hat{q}_k)$ can be ignored. The assumption of isotropic noise can make this filter, and the ones mentioned above, exceptionally robust to initial estimate errors and enhance their accuracy and convergence properties. However, since the aim here is to provide a general robust filter, we do not take this assumption.  

The measurement model \eqref{true_meas} can be reformulated similarly in the inertial frame where the noiseless accelerometer and magnetometer measurements should be $g$ and $b$ respectively, requiring us to rotate the actual sensor readings into the inertial frame. Instead of constructing the measurement model explicitly, we formulate directly the innovation $r_k$ as
\begin{align}
	\label{innov}
	r_k &= \B\left( \hat{q}_k \right) \bar{z} - \begin{bmatrix} g \\ b \end{bmatrix} = \begin{bmatrix} \A\left( \hat{q} \right) \bar{a}_k \\ \A\left( \hat{q} \right) \bar{m}_k \end{bmatrix} - \begin{bmatrix} g \\ b \end{bmatrix} ,  \\
	    &= \begin{bmatrix} \A\left( \hat{q}_k \right) \A\left( q_k \right)^T g \\ \A\left( \hat{q}_k \right) \A\left( q_k \right)^T b \end{bmatrix} + \begin{bmatrix} \A\left( \hat{q}_k \right) \nu^a_k \\ \A\left( \hat{q}_k \right) \nu^m_k \end{bmatrix} - \begin{bmatrix} g \\ b \end{bmatrix} , \nonumber \\
	    &= \begin{bmatrix} \left( \A(\Exp(\xi_k / 2)) - 1 \right) g \\ \left( \A(\Exp(\xi_k / 2)) - 1 \right) b \end{bmatrix} + \hat{\nu}_k, \nonumber \\
	    &\approx \begin{bmatrix} [\xi_k]_\times g \\ [\xi_k]_\times b \end{bmatrix} + \hat{\nu}_k, \nonumber \\
	\label{innov_linear}
	    &= \begin{bmatrix} \left[g\right]_\times \\ \left[b\right]_\times \end{bmatrix}\left( - \xi_k \right) + \hat{\nu}_k = H_k\left( -\xi_k  \right) + \hat{\nu}_k
\end{align}
where $\hat{\nu}_k = \B\left( \hat{q}_k \right)\nu_k$ and $q,\hat{q},\xi$ are as defined before (the superscripts $^-$ and $^+$ for a priori and posteriori states respectively are omitted here, since the equations \eqref{xi_prop}, \eqref{innov} and \eqref{innov_linear} are meant to be a general description of the propagation and measurement models). 

With $\xi$ as the state vector and $P$ denoting its error covariance, we now write the RI-EKF propagation and update equations as
\begin{align}
	\label{riekf_state_prop}
	\hat{q}_k^- &= \Phi_{k-1}\hat{q}_{k-1}^+, \qquad \xi_k^- = F_{k-1}\xi_{k-1}^+, \\
	\label{riekf_cov_prop}
	P_k^- &= F_{k-1}P_{k-1}^+F_{k-1} + \hat{Q}_k, \\
	\label{riekf_gain}
	K_k &= P_k^-H_k^T\left( H_kP_k^-H_k^T + \hat{R}_k \right), \\
	\label{riekf_state_update}
	\hat{q}_k^+ &= \Exp\left( \frac{K_k r_k}2 \right)\hat{q}_k^-, \qquad \xi_k^+ = \xi_k^- + K_k r_k, \\
	\label{riekf_cov_update}
	P_k^+ &= \left( I - K_kH_k \right) P_k^-,
\end{align}
where $F_k, r_k$ and $H_k$ are given by \eqref{xi_prop}, \eqref{innov} and \eqref{innov_linear} respectively and repeated below for convenience, along with the expressions for $\hat{Q}_k$ and $\hat{R}_k$:
\begin{align}
	F_k &= I_3,  \quad r_k = B\left( \hat{q}_k^- \right) \bar{z}_k - \begin{bmatrix} g \\ b \end{bmatrix} ,\quad H_k= \begin{bmatrix} \left[g\right]_\times \\ \left[b\right]_\times \end{bmatrix}, \nonumber \\
	\label{QR_def}
	\hat{Q}_k &=  \A\left(\hat{q}_k^-\right)\Sigma_\eta\left(\Delta t\right)^2 \A\left(\hat{q}^-_k\right)^T,  \hat{R}_k = \B\left( \hat{q}_k^- \right) \Sigma_\nu \B\left( \hat{q}_k^- \right)^T 
,\end{align}
and $\Phi$ is obtained from discretizing equation \eqref{noiseless_prop}:
\begin{equation*}
    \Phi_k = \exp\left\{\frac{\Delta t}{2}\Omega\left[\omega_k\right]\right\},
\end{equation*}
where $\exp$ is the matrix exponential, given for a matrix $A$ by
\[\exp{A} = \sum_{k=0}^\infty \frac{A^k}{k!}.\]

\subsection{Noise Covariance Estimation} \label{sec:noise_est}

We extend the expectation-maximization (EM) approach proposed in \cite{shumway1982} to accommodate the time-variant process and the non-linear measurement models. Further, the parameters to be estimated are no longer $Q,R$ but rather $\Sigma_\eta,\Sigma_\nu$, related by \eqref{QR_def}. We write the set of a parameters $\Theta$ as $\left\{ \mu_0, \Sigma_0, \Sigma_\eta, \Sigma_\nu\right\}$ where  $\mu_0,\Sigma_0$ are the initial Kalman state and error covariance estimates. The log-likelihood function is formulated over a window of length $n$ as
\begin{align}
	\label{log_lik}
	-2&\ln{L\left(\Theta  \right)} \coloneq \nonumber\\
	  &\ln\lvert\Sigma_0\rvert + \left( \xi_0 - \mu_0 \right)\Sigma_0^{-1}\left( \xi_0 - \mu_0 \right)^T \nonumber\\
		 &+\sum_{i=1}^n\ln\lvert Q_i\rvert + \left( \xi_i - F_{i-1}\xi_{i-1} \right)^T Q_i^{-1}\left( \xi_i - F_{i-1}\xi_{i-1} \right) \nonumber\\
		 &+\sum_{i=1}^n \ln\lvert R_i\rvert + r_i^T R_i^{-1}r_i 
.\end{align}

We then write the conditional expectation of the log-likelihood function $G\left( \Theta \mid \Theta^j \right) \coloneq \E\left\{-2\ln{L\left( \Theta \right)} \mid \bar{z}_{1:n},\Theta^j\right\}$ in terms of the smoothed estimates $\xi^n_i, \hat{q}^n_i, P^n_i, P^n_{i,i-1}$ given by equations (\ref{rts_init}-\ref{lag1_iter}) in the Appendix:
\begin{align}
 \label{log_lik_expc}
 	G\left(\Theta \mid \Theta^{j}\right) &= \ln{\lvert \Sigma_0 \rvert} + n\ln\lvert \Sigma_\eta \rvert + n\ln\lvert \Sigma_\nu \rvert \nonumber \\
	&+ \tr\left\{\Sigma_0^{-1}\left[P^n_0 + \left( \xi^n_0 - \mu_0 \right) \left( \xi^n_0 - \mu_0 \right)^T\right]\right\} \nonumber \\
	&+ \tr\left\{\Sigma_\eta^{-1}\left[S_{11} - S_{10} - S_{10}^T + S_{00}\right]\right\}\left(\Delta t\right)^{-2} \nonumber \\
	&+ \tr\left\{\Sigma_\nu^{-1}\sum_{i=1}^nB_i^T\left[r^n_i r_i^{n^T} + H_iP^n_iH_i^T \right]B_i\right\}
,\end{align}
with $B_i \coloneq \B\left( \hat{q}^-_i \right),A_i \coloneq \A\left( \hat{q}^-_i \right)$ and
\begin{align*}
	r_i^n &= \B\left( \hat{q}^n_i \right)  \bar{z}_i - \begin{bmatrix} g^T&b^T \end{bmatrix}^T, \\
    S_{11} &= \sum_{i=1}^n A_i^T\left(\xi_i^n\xi_i^{n^T} + P_i^n\right)A_i, \\
    S_{10} &= \sum_{i=1}^nA_i^T\left(\xi_i^n\xi_{i-1}^{n^T} + P^n_{i,i-1}\right)F_{i-1}^TA_i, \\
    S_{00} &= \sum_{i=1}^nA_i^TF_{i-1}\left(\xi_{i-1}^n\xi_{i-1}^{n^T} + P^n_{i-1}\right)F_{i-1}^TA_i.
\end{align*}

Minimizing equation \eqref{log_lik_expc} with respect to the parameters $\Theta$ at iteration $j$, we get the estimates
\begin{align}
	\label{em_init}
	\mu^{j+1} &= \xi^n_0 \quad \text{and} \quad \Sigma_0^{j+1} = P^n_0 \\
	\label{em_Q}
	\Sigma_\eta^{j+1} &= \frac{1}{n}\left( S_{11} - S_{10}S_{00}^{-1}S_{10}^T \right)\left(\Delta t\right)^{-2} \\
	\label{em_R}
	\Sigma_\nu^{j+1} &= \frac{1}{n}\sum_{i=1}^nB_i^T\left[r^n_i r_i^{n^T} + H_iP^n_iH_i^T \right]B_i
\end{align}

Equations (\ref{log_lik_expc}-\ref{em_R}) are iterated over the same window length $n$ till either the log-likelihood function $G\left( \Theta\mid\Theta^j \right)$ or the parameters $\Theta^j$ converge. For a detailed discussion on the convergence properties of the EM algorithm, readers are directed to \cite{wu1983}. 
\section{Simulation Results}

We test the performance of the RI-EKF and the noise covariance estimation procedure over a simulated orientation trajectory. The performance of the RI-EKF and the LI-EKF without noise covariance estimation are compared by displaying the RI-EKF's convergence properties. Further, two sets of Monte Carlo simulation runs are conducted - the first with initial orientation errors and the second with incorrect noise covariance matrices and their subsequent estimation according to the EM algorithm described in Section \ref{sec:noise_est}.

\begin{figure}
    \centering
    \includegraphics{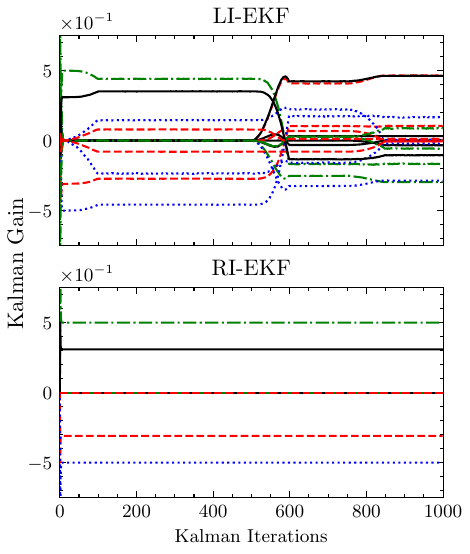}
    \caption{Elements of the Kalman gain matrix $K_k$ for the LI-EKF and RI-EKF.}
    \label{fig:gain}
\end{figure}

Figure \ref{fig:gain} compares the Kalman gain of the two filters over the same trajectory and initial state and error covariance estimates. The simulated noise is isotropic, with covariances $\Sigma_\eta = 10^{-1}I_3,  \Sigma_\nu = 10^{-5}I_6$. Note the variation in the LI-EKF gains. Its state transition matrix $F_k$ and measurement sensitivity matrix $H_k$ are both trajectory dependent \cite{gui2018,pandey2024}, leading to the variation of the gain with change in orientation. Meanwhile for the RI-EKF, equations \eqref{xi_prop} and \eqref{innov_linear}, along with the assumption of isotropic noise, ensure that the matrices $F_k$ and $H_k$, and hence the gain $K_k$, are independent of the trajectory.

\begin{figure}
    \centering
    \includegraphics{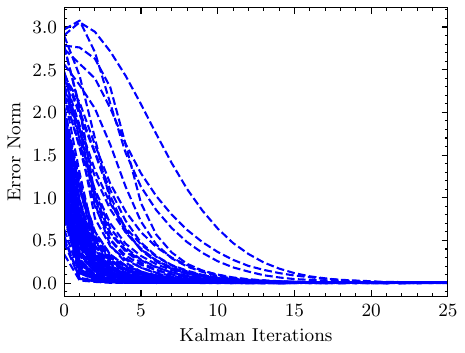}
    \caption{Convergence of Error Norm for the RI-EKF over different initial orientations.}
    \label{fig:err}
\end{figure}

The first set of 100 Monte Carlo runs are conducted with randomly generated initial orientations $q_0$ and a fixed initial estimate $\hat{q}^+_0 = \begin{bmatrix}
    1 & 0 & 0 & 0
\end{bmatrix}$. The initial estimation error is taken from the distribution $\xi^+_0\sim\mathcal{N}\left(0,I\right)$. Recall that $\xi^+_0$ is given by $\Exp\left(\xi^+_0/2\right)=\hat{q}^+_0\otimes q_0^{-1}.$
Figure \ref{fig:err} shows the convergence of the norm of the estimation error over iterations of the RI-EKF. 

\begin{figure}
    \centering
    \includegraphics{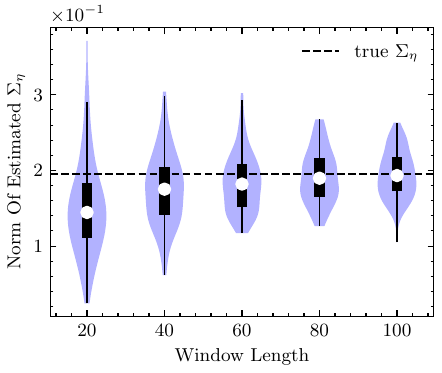}
    \caption{Gyroscope noise covariance estimation.}
    \label{fig:sigma_eta}
\end{figure}

\begin{figure}
    \centering
    \includegraphics{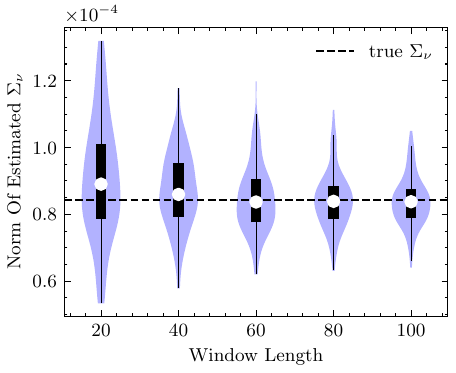}
    \caption{Accelerometer and magnetometer noise covariance estimation.}
    \label{fig:sigma_nu}
\end{figure}

Another set of 100 Monte Carlo runs are conducted to test the noise covariance estimation procedure with non-isotropic noise. The true covariance matrices are taken as $\Sigma_\eta^{true} = diag(\begin{bmatrix}
    0.75 & 1.5 & 1
\end{bmatrix})\times10^{-1}$ and $\Sigma_\nu^{true}=diag(\begin{bmatrix}
    1&2&3&3&3.5&6
\end{bmatrix})\times10^{-5}$ and the initial estimates as $\Theta^0 = \left\{400\Sigma_\eta^{true}, 200\Sigma_\nu^{true}\right\}.$ For each run, five RI-EKFs with differing window lengths - 20, 40, 60, 80, 100 - are iterated. Figures \ref{fig:sigma_eta} and \ref{fig:sigma_nu} show the distribution of the norm of the estimated covariance matrices as violin plots. We see that for each window length the estimates are concentrated around the true value, with optimal results for the window length 80.


\section{Conclusion}
In this article, we present an adaptive RI-EKF for quaternion-based attitude estimation, addressing varying sensor noise challenges while preserving geometric advantages. Our key findings include superior convergence properties of the RI-EKF compared to the LI-EKF, an effective noise covariance estimation procedure validated through Monte Carlo simulations. The method demonstrates robust performance across various window lengths, with optimal results at a window length of 80.  This approach shows significant potential for improving attitude estimation accuracy in dynamic, noisy environments, particularly in aerospace, robotics, and augmented and virtual reality applications.

\appendix[The Kalman Smoother] 

Let $\xi_k$ and $\bar{z}_k$ denote the state and measurement at time $k$ respectively. The a priori estimate $\xi_k^-$ can be formulated as the expected value of the state conditioned over measurements before time $k$. The posterior estimate $\xi_k^+$ can similarly be considered the expected value of the state conditioned over measurements upto (and including) time $k$:
\begin{align*}
	\xi^-_k&=\E\left\{\xi_k\mid \bar{z}_{1:k-1}\right\}, &&P^-_k = \var\left\{\xi_k\mid \bar{z}_{1:k-1}\right\} , \\ 
	\xi^+_k&=\E\left\{\xi_k\mid \bar{z}_{1:k}\right\}, &&P^+_k = \var\left\{\xi_k\mid \bar{z}_{1:k}\right\} 
.\end{align*}

In a similar vein, we can formulate the estimate conditioned to measurements obtained upto a time $n$ \textit{after} $k$, referred to as the smoothed estimate:
\begin{align*}
	\xi^n_k &\coloneq \E\left\{ \xi_k\mid \bar{z}_{1:n}\right\}, \qquad P^n_k \coloneq \var\left\{\xi_k\mid \bar{z}_{1:n}\right\} , \\ 
	P^n_{k,k-1} &\coloneq \var\left\{\xi_k,\xi_{k-1}\mid \bar{z}_{1:n}\right\}, \qquad \text{for} \quad k\le n
.\end{align*}  

The smoothed estimates are necessary for the computation of the log-likelihood function in equation \eqref{log_lik_expc}. $\xi^n_k$ and $P^n_k$ are determined iteratively by the Rauch-Tung-Striebel (RTS) smoother \cite{rauch1965} and $P^n_{k,k-1}$ is determined similarly by the lag-one covariance smoother \cite{shumway1982}. The smoothed quaternion $\hat{q}^n_k$ is also estimated using the smoothed state $\xi^n_k$.

We initialize the RTS smoother with
\begin{align}
    \label{rts_init}
    \xi^n_n = \xi_n^+,\quad
    \hat{q}^n_n = \hat{q}^+_n, \quad
    P_n = P_n^+,
\end{align}
and iterate for $i = n-1,\dots,1,0$
\begin{align}
    \label{rts_iter}
    J_i &= P_i^+F_i(P_{i+1}^-)^{-1}, \nonumber\\
    P_i^n &= P_i^+ + J_i(P_{i+1}^n - P_{i+1}^-)J_i^T, \nonumber\\
    \xi_i^n &= \xi_i^+ + J_i(\xi_{i+1}^n - \xi_{i+1}^-), \nonumber\\
    \hat{q}^n_i &= \Exp\left(\frac{\xi^n_i - \xi^+_i}2\right)\hat{q}^+_i 
\end{align}

Similarly, the lag-one covariance smoother is initialized as
\begin{align}
    \label{lag1_init}
    P^n_{n,n-1} &= (I-K_nH_n)F_{n-1}P_{n-1}^+,
\end{align}
and iterated for $i = n-1,\dots,1$
\begin{align}
    \label{lag1_iter}
    P^n_{i,i-1} &= P_i^+J_{i-1}^T + J_i(P^n_{i+1,i} - F_iP_i^+)J_{i-1}^T.
\end{align}

\bibliographystyle{IEEEtran}
\bibliography{IEEEabrv,acc}

\begin{thebibliography}{10}
\providecommand{\url}[1]{#1}
\csname url@samestyle\endcsname
\providecommand{\newblock}{\relax}
\providecommand{\bibinfo}[2]{#2}
\providecommand{\BIBentrySTDinterwordspacing}{\spaceskip=0pt\relax}
\providecommand{\BIBentryALTinterwordstretchfactor}{4}
\providecommand{\BIBentryALTinterwordspacing}{\spaceskip=\fontdimen2\font plus
\BIBentryALTinterwordstretchfactor\fontdimen3\font minus \fontdimen4\font\relax}
\providecommand{\BIBforeignlanguage}[2]{{%
\expandafter\ifx\csname l@#1\endcsname\relax
\typeout{** WARNING: IEEEtran.bst: No hyphenation pattern has been}%
\typeout{** loaded for the language `#1'. Using the pattern for}%
\typeout{** the default language instead.}%
\else
\language=\csname l@#1\endcsname
\fi
#2}}
\providecommand{\BIBdecl}{\relax}
\BIBdecl

\bibitem{schmidt1981}
S.~F. Schmidt, ``The {{Kalman}} filter - {{Its}} recognition and development for aerospace applications,'' \emph{Journal of Guidance and Control}, vol.~4, no.~1, pp. 4--7, 1981.

\bibitem{moore2016}
T.~Moore and D.~Stouch, ``A generalized extended kalman filter implementation for the robot operating system,'' in \emph{Intelligent Autonomous Systems 13: {{Proceedings}} of the 13th International Conference {{IAS-13}}}.\hskip 1em plus 0.5em minus 0.4em\relax Springer, 2016, pp. 335--348.

\bibitem{groves2015}
P.~D. Groves, ``Principles of {{GNSS}}, inertial, and multisensor integrated navigation systems, 2nd edition [{{Book}} review],'' \emph{IEEE Aerospace and Electronic Systems Magazine}, vol.~30, no.~2, pp. 26--27, Feb. 2015.

\bibitem{woodman2007}
O.~Woodman, ``An introduction to inertial navigation,'' 2007.

\bibitem{titterton2004}
D.~Titterton and J.~L. Weston, \emph{Strapdown {{Inertial Navigation Technology}}}.\hskip 1em plus 0.5em minus 0.4em\relax IET, 2004.

\bibitem{quinchia2013}
A.~G. Quinchia, G.~Falco, E.~Falletti, F.~Dovis, and C.~Ferrer, ``A {{Comparison}} between {{Different Error Modeling}} of {{MEMS Applied}} to {{GPS}}/{{INS Integrated Systems}},'' \emph{Sensors}, vol.~13, no.~8, pp. 9549--9588, Aug. 2013.

\bibitem{kalman1960}
R.~E. Kalman, ``A new approach to linear filtering and prediction problems,'' \emph{Transactions of the ASME--Journal of Basic Engineering}, vol.~82, no. Series D, pp. 35--45, 1960.

\bibitem{crassidis2007}
J.~L. Crassidis, F.~L. Markley, and Y.~Cheng, ``Survey of {{Nonlinear Attitude Estimation Methods}},'' \emph{Journal of Guidance, Control, and Dynamics}, vol.~30, no.~1, pp. 12--28, Jan. 2007.

\bibitem{simon_ekf}
D.~Simon, ``Nonlinear kalman filtering,'' in \emph{Optimal State Estimation}.\hskip 1em plus 0.5em minus 0.4em\relax John Wiley \& Sons, Ltd, 2006, ch.~13, pp. 393--431.

\bibitem{bonnabel2009}
S.~Bonnabel, P.~Martin, and E.~Salaun, ``Invariant {{Extended Kalman Filter}}: Theory and application to a velocity-aided attitude estimation problem,'' in \emph{Proceedings of the 48h {{IEEE Conference}} on {{Decision}} and {{Control}} ({{CDC}}) Held Jointly with 2009 28th {{Chinese Control Conference}}}.\hskip 1em plus 0.5em minus 0.4em\relax Shanghai: IEEE, Dec. 2009, pp. 1297--1304.

\bibitem{gui2018}
H.~Gui and A.~H.~J. De~Ruiter, ``Quaternion {{Invariant Extended Kalman Filtering}} for {{Spacecraft Attitude Estimation}},'' \emph{Journal of Guidance, Control, and Dynamics}, vol.~41, no.~4, pp. 863--878, Apr. 2018.

\bibitem{barrau2018}
A.~Barrau and S.~Bonnabel, ``Invariant {{Kalman Filtering}},'' \emph{Annual Review of Control, Robotics, and Autonomous Systems}, vol.~1, no.~1, pp. 237--257, May 2018.

\bibitem{barrau2023}
------, ``The {{Geometry}} of {{Navigation Problems}},'' \emph{IEEE Transactions on Automatic Control}, vol.~68, no.~2, pp. 689--704, Feb. 2023.

\bibitem{mehra1970}
R.~Mehra, ``On the identification of variances and adaptive {{Kalman}} filtering,'' \emph{IEEE Transactions on Automatic Control}, vol.~15, no.~2, pp. 175--184, 1970.

\bibitem{mohamed1999}
A.~H. Mohamed and K.~P. Schwarz, ``Adaptive kalman filtering for {{INS}}/{{GPS}},'' \emph{Journal of Geodesy}, vol.~73, no.~4, pp. 193--203, May 1999.

\bibitem{shumway1982}
R.~H. Shumway and D.~S. Stoffer, ``An approach to time series smoothing and forecasting using the {{EM}} algorithm,'' \emph{Journal of Time Series Analysis}, vol.~3, no.~4, pp. 253--264, Jul. 1982.

\bibitem{ananthasayanam2016}
M.~R. Ananthasayanam, M.~S. Mohan, N.~Naik, and R.~M.~O. Gemson, ``A heuristic reference recursive recipe for adaptively tuning the {{Kalman}} filter statistics part-1: Formulation and simulation studies,'' \emph{S{\=a}dhan{\=a}}, vol.~41, no.~12, pp. 1473--1490, Dec. 2016.

\bibitem{shoemake1985}
\BIBentryALTinterwordspacing
K.~Shoemake, ``Animating rotation with quaternion curves,'' in \emph{Proceedings of the 12th Annual Conference on Computer Graphics and Interactive Techniques}, ser. SIGGRAPH '85.\hskip 1em plus 0.5em minus 0.4em\relax New York, NY, USA: Association for Computing Machinery, 1985, p. 245–254. [Online]. Available: \url{https://doi.org/10.1145/325334.325242}
\BIBentrySTDinterwordspacing

\bibitem{sola2017}
J.~Sol{\`a}, ``Quaternion kinematics for the error-state {{Kalman}} filter,'' \emph{CoRR}, vol. abs/1711.02508, 2017.

\bibitem{pandey2024}
Y.~Pandey, R.~Bhattacharyya, and Y.~N. Singh, ``Robust attitude estimation with quaternion left-invariant {{EKF}} and noise covariance tuning,'' \emph{arXiv preprint arXiv:2409.11496}, 2024.

\bibitem{markley2004}
F.~L. Markley, ``Attitude estimation or quaternion estimation?'' \emph{The Journal of the Astronautical Sciences}, vol.~52, pp. 221--238, 2004.

\bibitem{bourmaud2015}
G.~Bourmaud, R.~M{\'e}gret, M.~Arnaudon, and A.~Giremus, ``Continuous-discrete extended kalman filter on matrix lie groups using concentrated gaussian distributions,'' \emph{Journal of Mathematical Imaging and Vision}, vol.~51, no.~1, 2015.

\bibitem{bourmaud2013}
G.~Bourmaud, R.~M{\'e}gret, A.~Giremus, and Y.~Berthoumieu, ``Discrete extended kalman filter on lie groups,'' in \emph{21st European Signal Processing Conference ({{EUSIPCO}} 2013)}, 2013, pp. 1--5.

\bibitem{hauberg2013}
S.~Hauberg, F.~Lauze, and K.~S. Pedersen, ``Unscented kalman filtering on riemannian manifolds,'' \emph{Journal of Mathematical Imaging and Vision}, vol.~46, pp. 103--120, 2013.

\bibitem{hartley2020}
R.~Hartley, M.~Ghaffari, R.~M. Eustice, and J.~W. Grizzle, ``Contact-aided invariant extended {{Kalman}} filtering for robot state estimation,'' \emph{The International Journal of Robotics Research}, vol.~39, no.~4, pp. 402--430, Mar. 2020.

\bibitem{barrau2016}
A.~Barrau and S.~Bonnabel, ``The invariant extended {{Kalman}} filter as a stable observer,'' \emph{IEEE Transactions on Automatic Control}, vol.~62, no.~4, pp. 1797--1812, 2016.

\bibitem{wu1983}
C.~F.~J. Wu, ``On the convergence properties of the {{EM}} algorithm,'' \emph{The Annals of Statistics}, vol.~11, no.~1, pp. 95--103, 1983.

\bibitem{rauch1965}
H.~E. Rauch, F.~Tung, and C.~T. Striebel, ``Maximum likelihood estimates of linear dynamic systems,'' \emph{AIAA Journal}, vol.~3, no.~8, pp. 1445--1450, 1965.

\end{thebibliography}

\end{document}